\documentclass[aps,prl,floatfix,onecolumn]{revtex4}%
\usepackage{amssymb}
\usepackage{amsmath}
\usepackage{graphicx}
\usepackage{amsfonts}%
\providecommand{\U}[1]{\protect\rule{.1in}{.1in}}
\providecommand{\U}[1]{\protect\rule{.1in}{.1in}}
\providecommand{\U}[1]{\protect\rule{.1in}{.1in}}
\providecommand{\U}[1]{\protect\rule{.1in}{.1in}}
\providecommand{\U}[1]{\protect\rule{.1in}{.1in}}
\providecommand{\U}[1]{\protect\rule{.1in}{.1in}}
\providecommand{\U}[1]{\protect\rule{.1in}{.1in}}
\providecommand{\U}[1]{\protect\rule{.1in}{.1in}}
\providecommand{\U}[1]{\protect\rule{.1in}{.1in}}
\providecommand{\U}[1]{\protect\rule{.1in}{.1in}}
\begin{document}
\title{Subradiant split Cooper pairs}
\author{Audrey Cottet$^{1}$ , Takis Kontos$^{1}$ and Alfredo Levy Yeyati$^{2}$}
\affiliation{$^{1}$Laboratoire Pierre Aigrain, Ecole Normale Sup\'{e}rieure, CNRS UMR 8551,
Laboratoire associ\'{e} aux universit\'{e}s Pierre et Marie Curie et Denis
Diderot, 24, rue Lhomond, 75231 Paris Cedex 05, France}
\affiliation{$^{2}$Departamento de F\'{\i}sica Te\'{o}rica de la Materia Condensada C-V and
Instituto Nicol\'{a}s Cabrera, Universidad Aut\'{o}noma de Madrid, E-28049
Madrid, Spain}

\begin{abstract}
We suggest a way to characterize the coherence of the split Cooper pairs
emitted by a double-quantum-dot based Cooper pair splitter (CPS), by studying
the radiative response of such a CPS inside a microwave cavity. The coherence
of the split pairs manifests in a strongly nonmonotonic variation of the
emitted radiation as a function of the parameters controlling the coupling of
the CPS to the cavity. The idea to probe the coherence of the electronic
states using the tools of Cavity Quantum Electrodynamics could be generalized
to many other nanoscale circuits.

\end{abstract}
\maketitle

Entanglement is now accepted as an intriguing but available resource of the
quantum world. However, it is still unclear whether this phenomenon can
survive in nanoscale electronic circuits, because an electronic fluid is
characterized by a complex many-body state in general. Producing and detecting
entangled electronic states are therefore important goals of quantum
electronics. The spin-singlet (Cooper) pairing of electrons, naturally present
in conventional superconductors, appears as a very attractive source of
electronic entangled states. Recently, the splitting of Cooper pairs could be
demonstrated in Y-junctions made out of nanowires \cite{Herrmann:10,
Hofstetter:09,Hofstetter:11}. But how to distinguish between a singlet state
broken into a product state due to decoherence and the desired coherent
singlet state is not addressed in these experiments. Here, we show that the
radiative response of such devices can reveal the presence or absence of
entangled states. The split Cooper pairs are shown to decouple from the
electromagnetic field conveyed by a photonic cavity (subradiance) if coherent.
These findings add a new twist to quantum opto-electronics, and could be
applied to any source of on-demand entangled electronic
states.\begin{figure}[h]
\includegraphics[width=0.5\linewidth]{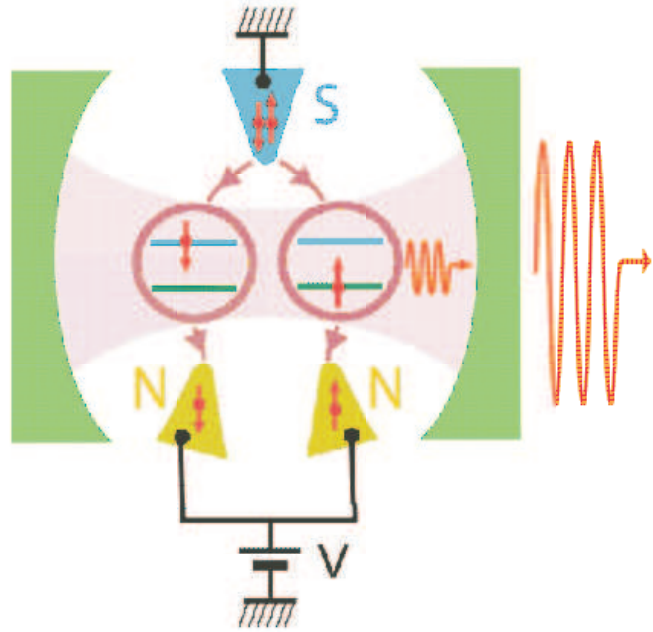}\caption{Scheme of the CPS
embedded in a \ photonic cavity. The CPS is made out of a double quantum dot
coupled to a central blue superconducting contact ($S$) connected to ground,
and two outer yellow normal metal contacts (N) biased with a voltage $V$. This
circuit is placed inside a photonic cavity (represented schematically by
mirrors in green). The Cooper pairs spread over the $K$ and $K^{\prime}$
orbitals of the dots, sketched in blue and green respectively. The low energy
level structure of the system allows photon emission which can be amplified
through a lasing effect.}%
\end{figure}

The analogy between a beam splitter for electronic states and a beam splitter
for photons calls for the use of correlation measurements to characterize the
degree of entanglement of pairs of electrons, via the noise cross-correlations
of the electrical current
\cite{Martin:96,Anantram,Burkard:99,Recher:01,Lesovik,Borlin:02,Samuelsson:02}%
. The use of a double quantum dot circuit connected to two normal electrodes
and one superconducting electrode gives a practical realization of an
electronic entangler and simplifies the diagnosis of entanglement from
transport, as originally suggested by Recher et al. \cite{Recher:01}.
Nevertheless, measuring cross-correlations in such a setup is a formidable
task because the conditions for useful entanglement correspond to a regime
where the system is almost isolated from the leads. In this case, the Cooper
pair current is too small to yield measurable current fluctuations with
present amplification techniques \cite{Wei}. These difficulties stem from the
fact that probing directly a quantum system with transport measurements is not
natural since one needs the system to be open and closed at the same time. On
the contrary, as known in atomic physics \cite{Raimond:01}, the use of
light-matter interaction is very well adapted to probe closed quantum systems.
Here, we show how to use the coupling between electrons and photons from a
microwave cavity, to assess the coherence of Cooper pairs emitted by a CPS
implemented with a double quantum dot circuit.

We consider a CPS made out of a single wall carbon nanotube in which a double
dot is defined by a central superconducting contact connected to ground and
two outer normal metal contacts biased with a voltage $V$. This CPS is
inserted inside a photonic cavity (see Figure 1) which is assumed to be
implemented in a coplanar waveguide geometry using a superconducting metal,
like in the Circuit Quantum Electrodynamics architecture \cite{Wallraff:04}.
Very recently, it has been demonstrated experimentally \cite{Delbecq,Frey}
that one can extend this architecture to quantum dot circuits
\cite{Childress,Cottet2010,Cottet2011,Jin:11}, bringing our proposed setup
within experimental reach.\begin{figure}[h]
\includegraphics[width=0.55\linewidth]{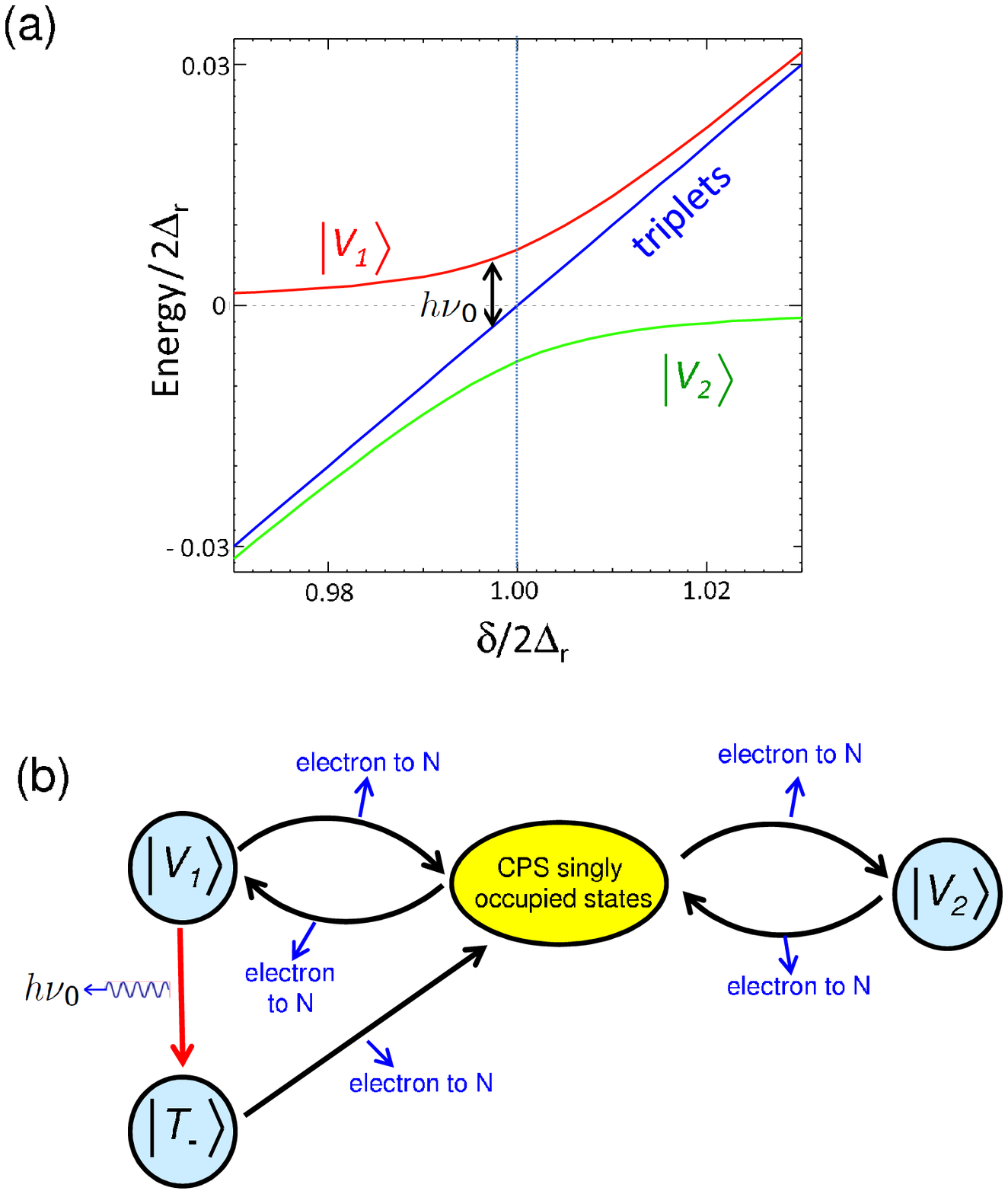}\caption{a. Spectrum of the
subset $\mathcal{E}$ of even charge states of the CPS near $\delta=2\Delta
_{r}$. It comprizes three spin-triplet states, and two states $\left\vert
V_{1}\right\rangle $ and $\left\vert V_{2}\right\rangle $ which are a coherent
mixture of the empty state and a spin-singlet state $\left\vert \mathcal{S}%
\right\rangle $. b. Dynamics of the CPS near the working point $\delta
=2\Delta_{r}$. We consider a bias voltage regime such that the various
transitions between the CPS states occur together with the transfer of one
electron towards the normal contacts, apart from the transition from
$\left\vert V_{1}\right\rangle $\ to the triplet state $\left\vert
T_{-}\right\rangle $, which occurs due to the emission of a photon towards the
cavity. We have checked that the radiative transition from $\left\vert
T_{-}\right\rangle $ to $\left\vert V_{2}\right\rangle $ is not active in the
regime of parameters we consider.}%
\end{figure}

Our aim is to find an effect which reveals qualitatively the coherent
injection of Cooper pairs. Thanks to Coulomb blockade, the states
participating to electronic transport can be reduced to the double-dot empty
state $\left\vert 0,0\right\rangle $, the singly occupied states $\left\vert
\tau\sigma,0\right\rangle =d_{L\tau\sigma}^{\dag}\left\vert 0,0\right\rangle $
and $\left\vert 0,\tau\sigma\right\rangle =d_{R\tau\sigma}^{\dag}\left\vert
0,0\right\rangle $ of the left (L) and right (R) dot, and the non-local doubly
occupied states $\left\vert \tau\sigma,\tau^{\prime}\sigma^{\prime
}\right\rangle =d_{L\tau\sigma}^{\dag}d_{R\tau^{\prime}\sigma^{\prime}}^{\dag
}\left\vert 0,0\right\rangle $, with $\tau\in\{K,K^{\prime}\}$ and $\sigma
\in\{\uparrow,\downarrow\}$ being orbital and spin indices respectively (see
Supplemental Material (SM), part A for details). For biasing conditions like
the one depicted in Fig. 1 with $eV<\Delta$ ($\Delta$ being the gap of the
superconducting contact) Cooper pairs are injected from the superconductor
into a singlet state $\left\vert \mathcal{S}\right\rangle $. The coupling to
the superconductor hybridizes the $\left\vert \mathcal{S}\right\rangle $ state
with the empty state $\left\vert 0,0\right\rangle $ forming two states which
we call $\left\vert V_{1}\right\rangle $ and $\left\vert V_{2}\right\rangle $.
These states can relax to a single particle state via electron tunneling into
the normal leads, and then via a second tunneling process into the state
$\left\vert V_{1}\right\rangle $ or $\left\vert V_{2}\right\rangle $, which
closes the usual operation loop of the CPS, as depicted in Fig. 2b. However,
when the CPS is placed in the resonant cavity there is an additional
transition to a triplet state (two electronic states with equal spin) through
the emission of a cavity photon. It is the aim of this work to study these
photon emission processes and show that they can be used to characterize the
coherence of the split pairs injected into the CPS.

The relevant energy scales of the CPS are the position $\varepsilon$ of the
energy levels on each dot, which we assume symmetric for the sake of
simplicity, the Cooper pair coherent splitting rate $t_{eh}$
\cite{Eldridge,LY}, the effective spin-orbit coupling constant $\Delta_{so}$
and the coupling between the $K$ and $K^{\prime}$ orbitals, $\Delta
_{K\leftrightarrow K^{\prime}}$, which arise from weak disorder in the
nanotube \cite{Jespersen,Liang,Kuemmeth,Palyi}. For $t_{eh}=0$, $\Delta
_{so}=0$, and $\Delta_{K\leftrightarrow K^{\prime}}=0$, the doubly occupied
states cost an energy $2\varepsilon=\delta$. For $t_{eh}\ll\Delta
_{K\leftrightarrow K^{\prime}},\Delta_{so}$ \cite{Herrmann:10}, the regime
$\delta\sim2\Delta_{r}$, with $\Delta_{r}=\sqrt{\Delta_{so}^{2}+\Delta
_{K\leftrightarrow K^{\prime}}^{2}}$, is the most adequate for producing
entangled states. It allows one to isolate, in the double dot even charge
sector, a subset $\mathcal{E}$ of five lowest energy eigenstates which are at
a distance $\sim2\Delta_{r}$ from the other even charge states, at least.
These five states include three spin-triplet states $\left\vert T_{0}%
\right\rangle $, $\left\vert T_{+}\right\rangle $, and $\left\vert
T_{-}\right\rangle $ with energy $E_{triplet}=\delta-2\Delta_{r}$ and the two
hybridized even states \cite{Recher:11,Godschalk} $\left\vert V_{n}%
\right\rangle =\sqrt{1-v_{n}^{2}}\left\vert 0,0\right\rangle +v_{n}\left\vert
\mathcal{S}\right\rangle $ with energy $E_{i}=\frac{1}{2}\left(
\delta-2\Delta_{r}-(-1)^{n}\sqrt{8t_{eh}^{2}+(\delta-2\Delta_{r})^{2}}\right)
$ for $n\in\{1,2\}$ (see SM, part B.4, for the expression of the
$v_{n}^{\prime}s$). Here, the definition of the spin-singlet state $\left\vert
\mathcal{S}\right\rangle $ and the spin-triplets take into account the twofold
orbital degeneracy of each dot (see SM, part B.4).\textbf{ }The energies of
$\left\vert V_{1}\right\rangle $ and $\left\vert V_{2}\right\rangle $ show an
anticrossing with a width $E_{p}=2\sqrt{2}t_{eh}$ at $\delta=2\Delta_{r}$,
while the energy of the triplets lies in the middle (see Figure 2.a).

Let us first discuss the current $I_{CPS}$ through the superconducting lead,
without considering the effect of the photonic cavity. We call $\Gamma_{N}$
the bare tunnel rate of an electron between one dot and the corresponding
normal metal contact, while $P_{A}$ denotes the probability of a state
$\left\vert A\right\rangle $ of the double dot even charge sector and
$P_{\text{single}}$ the global probability of having a double dot singly
occupied state. In the sequential tunneling limit $\Gamma_{N}\ll k_{B}T$,
these probabilities follow a master equation $\frac{d}{dt}P=M$ $P$ with%
\begin{equation}
P=\left[
\begin{array}
[c]{c}%
P_{V_{1}}\\
P_{V_{2}}\\
P_{T_{-}}\\
P_{\text{single}}%
\end{array}
\right]  \text{ , }\frac{M}{\Gamma_{N}}=\left[
\begin{array}
[c]{cccc}%
-2v_{1}^{2} & 0 & 0 & 1-v_{1}^{2}\\
0 & -2v_{2}^{2} & 0 & 1-v_{2}^{2}\\
0 & 0 & -2 & 0\\
2v_{1}^{2} & 2v_{2}^{2} & 2 & -1
\end{array}
\right]  \label{master}%
\end{equation}
Equation (\ref{master}) corresponds to a bias voltage regime such that single
electrons can go from the double dot to the normal leads but not the reverse
(see SM, part C). We disregard the states $\left\vert T_{0}\right\rangle $ and
$\left\vert T_{+}\right\rangle $, which are not populated in the simple limit
we consider. The states $\left\vert V_{i}\right\rangle $ and $\left\vert
T_{-}\right\rangle $ can decay towards several different singly occupied
states. The sum of the corresponding transition rates equals $2\Gamma_{N}%
v_{i}^{2}$ and $2\Gamma_{N}$ respectively, which explains the presence of the
factors 2 in the 3 first columns of $M$. To calculate $I_{CPS}$, one first has
to determine the stationary value $P_{stat}$ of $P$\ from $MP_{stat}=0$. Then,
one can use $I_{CPS}=R.P_{stat}$ with $R=e\Gamma_{N}[2v_{1}^{2},2v_{2}%
^{2},2,1]$. If the double dot is initially in the state $\left\vert
V_{n}\right\rangle $, with $n\in\{1,2\}$, there can be a transition to a
singly occupied state while an electron is transferred to the normal leads,
because $\left\vert V_{n}\right\rangle $ has a $\left\vert \mathcal{S}%
\right\rangle $\ component. Then, there can be a transition from this singly
occupied state to $\left\vert V_{m}\right\rangle $, with $m\in\{1,2\}$,
because $\left\vert V_{m}\right\rangle $ has a $\left\vert 0,0\right\rangle
$\ component. This leads to the existence of state cycles which produce a flow
of electrons towards the normal leads (see Figure 2.b). On the left(right) of
the anticrossing point $\delta=2\Delta_{r}$, the state $\left\vert
V_{1(2)}\right\rangle $ is the most probable. Exactly at the anticrossing
point, the states $\left\vert V_{1}\right\rangle $ and $\left\vert
V_{2}\right\rangle $ contribute equally to current transport and $I_{CPS}$ is
maximum (see Fig.3, main frame).\begin{figure}[h]
\includegraphics[width=0.55\linewidth]{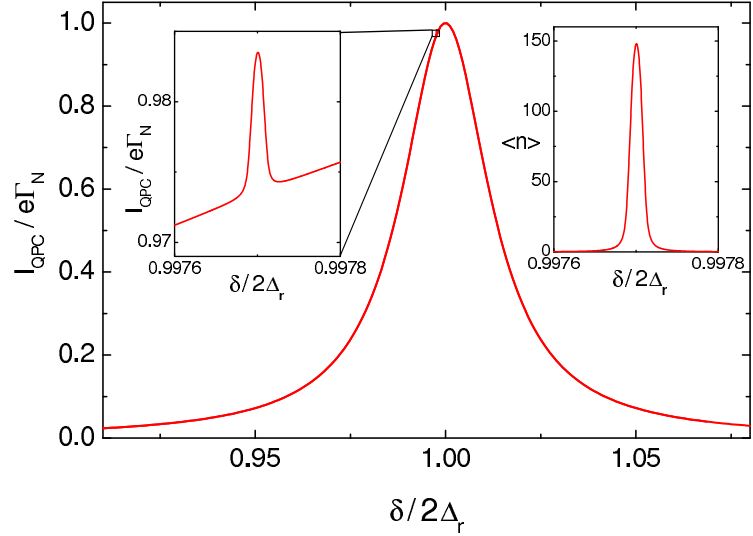}\caption{Main frame:
Current $I_{QPC}$ at the input of the Cooper pair splitter. The peak at
$\delta=2\Delta_{r}$ is due to the anticrossing between $\left\vert
V_{1}\right\rangle $ and $\left\vert V_{2}\right\rangle $. Left inset: zoom on
the current peak due to lasing transitions between the $\left\vert
V_{1}\right\rangle $ and $\left\vert T_{-}\right\rangle $ states. Right inset:
number of photons versus $\delta$. We have used realistic parameters
$t_{eh}=9~\mathrm{\mu eV}$, $\Delta_{so}=0.15~\mathrm{meV}$, $\Delta
_{K/K^{\prime}}=0.9~\mathrm{meV}$, $\nu_{0}=3.64~\mathrm{GHz}$, $Q=2\pi
\hbar\nu_{0}/\kappa=500000$, $D_{pq}^{-1}=92~\mathrm{ns}$, $\Gamma_{N}=72$
\textrm{M}$\mathrm{Hz}$, $T=35~\mathrm{mK}$, and $\left\vert \lambda
_{L}-\lambda_{R}\right\vert =2.5$ $10^{-6}\Delta_{r}\simeq0.55~\mathrm{MHz}$.}%
\end{figure}

Electronic spins are naturally coupled to photons thanks to the spin-orbit
interaction, which exists in many types of conductors, including those used to
demonstrate Cooper pair splitting. In our case, we will take into account a
spin/photon coupling with the form (see SM, part F)%
\begin{equation}
H_{so}=(a+a^{\dag})%
{\displaystyle\sum\limits_{i,\tau,\sigma}}
\lambda_{i\sigma}d_{i\tau\sigma}^{\dag}d_{i\tau\overline{\sigma}}=(a+a^{\dag
})h_{so} \label{hso}%
\end{equation}
with $d_{i\tau\sigma}^{\dag}$ the creation operator for an electron with spin
$\sigma$ in orbital $\tau$ of dot $i\in\{L,R\}$ and $a^{\dag}$ the creation
operator for cavity photons. We use $\lambda_{i\sigma}=\mathbf{i}\sigma
\lambda_{i}$. Inside the $\mathcal{E}$ subspace, the term $H_{so}$ couples
$\left\vert V_{n}\right\rangle $ to $\left\vert T_{_{-}}\right\rangle $ only,
i.e., for $n\in\{1,2\}$,
\begin{equation}
\left\langle T_{_{-}}\right\vert h_{so}\left\vert V_{n}\right\rangle
=\mathbf{i}v_{n}\frac{\Delta_{K\leftrightarrow K^{\prime}}}{\Delta_{r}%
}(\lambda_{L}-\lambda_{R}) \label{alpha1}%
\end{equation}
whereas $\left\langle T_{+[0]}\right\vert h_{so}\left\vert V_{1(2)}%
\right\rangle =0$. The presence of the minus sign in Eq.(\ref{alpha1}) is
crucial. It reveals the \textit{coherent} injection of singlet pairs inside
the CPS. If $\left\vert V_{n}\right\rangle $ was resulting from the
hybridization of a product spin state with the $\left\vert 0,0\right\rangle $
state, the matrix element (\ref{alpha1}) would depend only on $\lambda_{L}$ or
$\lambda_{R}$. We will show below that the peculiar structure of the element
(\ref{alpha1}) can be revealed by measuring the lasing effect associated to
the transition $\left\vert V_{1}\right\rangle \rightarrow\left\vert T_{_{-}%
}\right\rangle $.

In order to study current transport and the photonic dynamics simultaneously,
one can generalize the above master equation description by using a
semi-quantum description of lasing \cite{Andre}. This consists in adding
inside the matrix $M$ rates accounting for photonic emission and absorption
processes, i.e. $M\rightarrow M+M_{ph}$ with%
\begin{align}
M_{ph} &  =\left[
\begin{tabular}
[c]{llll}%
$0$ & $0$ & $W_{V_{1}T_{-}}$ & $0$\\
$0$ & $-W_{T_{-}V_{2}}$ & $0$ & $0$\\
$0$ & $W_{T_{-}V_{2}}$ & $-W_{V_{1}T_{-}}$ & $0$\\
$0$ & $0$ & $0$ & $0$%
\end{tabular}
\ \ \ \ \ \ \ \ \ \ \ \right]  \left\langle n\right\rangle \nonumber\\
&  +\left[
\begin{tabular}
[c]{llll}%
$-W_{V_{1}T_{-}}$ & $0$ & $0$ & $0$\\
$0$ & $0$ & $W_{T_{-}V_{2}}$ & $0$\\
$W_{V_{1}T_{-}}$ & $0$ & $-W_{T_{-}V_{2}}$ & $0$\\
$0$ & $0$ & $0$ & $0$%
\end{tabular}
\ \ \ \ \ \ \ \ \ \ \ \right]  \left(  \left\langle n\right\rangle +1\right)
\end{align}
and $\left\langle n\right\rangle $ the average number of photons in the
cavity. The master equation must be solved self-consistently with a photonic
balance equation%
\begin{align}
0 &  =P_{V_{1}}W_{V_{1}T_{-}}\left(  \left\langle n\right\rangle +1\right)
-P_{V_{2}}W_{T_{-}V_{2}}\left\langle n\right\rangle \label{P}\\
&  +P_{T_{-}}[W_{T_{-}V_{2}}\left(  \left\langle n\right\rangle +1\right)
-W_{V_{1}T_{-}}\left\langle n\right\rangle ]-\kappa(\left\langle
n\right\rangle -\left\langle n\right\rangle _{th})\nonumber
\end{align}
with $\left\langle n\right\rangle _{th}=\left(  \exp(\hbar\omega_{0}%
/k_{B}T)-1\right)  ^{-1}$. The photonic transition rates $W_{V_{1}T_{-}}$ and
$W_{T_{-}V_{2}}$ appearing in the above equations can be expressed as:%
\begin{equation}
W_{pq}=\frac{2\left\vert \left\langle p\right\vert h_{so}\left\vert
q\right\rangle \right\vert ^{2}\left(  \frac{\kappa}{2}+D_{qp}\right)
}{(E_{p}-E_{q}-2\pi\hbar\nu_{0})^{2}+\left(  \frac{\kappa}{2}+D_{qp}\right)
^{2}}%
\end{equation}
where $\kappa$ is the damping rate of the resonator and $D_{qp}$ the
decoherence rate associated to the $p\longleftrightarrow q$ resonance. The
above description is valid provided the transitions $\left\vert V_{1}%
\right\rangle \rightarrow\left\vert T_{_{-}}\right\rangle $ and $\left\vert
T_{_{-}}\right\rangle \rightarrow\left\vert V_{2}\right\rangle $ are not both
resonant with the cavity, i.e. one does not have simultaneously $\delta
=2\Delta_{r}$ and $2\pi\hbar\nu_{0}=\sqrt{2}t_{eh}$. The left inset of Figure
3 shows a zoom on $I_{CPS}$ around $\delta=\delta_{l}(\nu_{0})=2\Delta
_{r}-2\pi\hbar\nu_{0}+(t_{eh}^{2}/\pi\hbar\nu_{0})$. A current peak occurs at
$\delta=\delta_{l}(\nu_{0})$, due to the lasing effect which involves the
$\left\vert V_{1}\right\rangle \rightarrow\left\vert T_{_{-}}\right\rangle $
transition. Note that the population inversion necessary for the lasing effect
is achieved without any AC excitation thanks to the DC bias voltage. The
tunnel transition rate from $\left\vert T_{_{-}}\right\rangle $ to the singly
occupied states is larger than the transition rate from $\left\vert
V_{1}\right\rangle $ to the singly occupied states (see Eq.(\ref{master})),
which explains the increase in $I_{CPS}$ while the system lases. However, for
typical parameters, the current peak due to lasing corresponds to
$\sim100~\mathrm{fA}$ over a strong background of $\sim10~\mathrm{pA}$. The
lasing effect is more clearly visible through the average number $\left\langle
n\right\rangle $ of photons in the cavity, which can be measured with
microwave amplification techniques \cite{Astafiev} (see right inset of figure
3).\begin{figure}[h]
\includegraphics[width=0.7\linewidth]{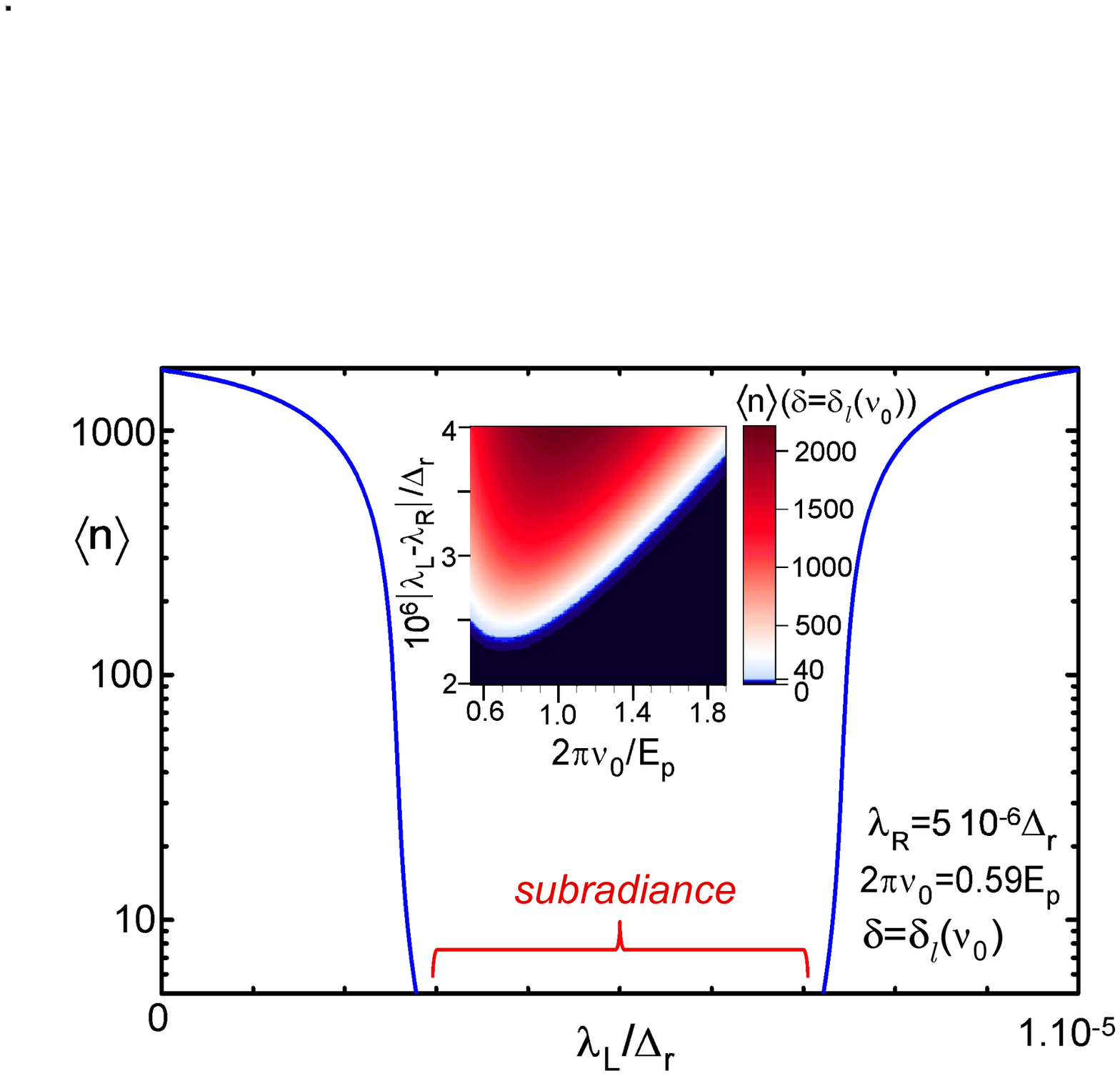}\caption{Main frame: number
$\left\langle n\right\rangle $ of photons in the cavity for the same
parameters as in Figure 3, as a function of the coupling $\lambda_{L}$, for
constant values of $\delta$ and $\lambda_{R}$. The vanishing of $\left\langle
n\right\rangle $ for small values of $\left\vert \lambda_{L}-\lambda
_{R}\right\vert $ is a direct signature of coherent injection of split Cooper
pairs inside the CPS. Inset: maximum number of photons which can be obtained
at $\delta=\delta_{l}(\nu_{0})$, as a function of the cavity frequency
$\nu_{0}$ and $\left\vert \lambda_{L}-\lambda_{R}\right\vert $. If $\left\vert
\lambda_{L}-\lambda_{R}\right\vert $ is sufficiently large, the lasing effect
can be obtained for a broad range of $\nu_{0}$.}%
\end{figure}

The peculiar shape of the matrix element (\ref{alpha1}) is particularly
interesting if one can tune independently the couplings $\lambda_{L(R)}$, e.g.
tune $\lambda_{L}$ while $\lambda_{R}$ remains constant (see SM, part F.2).
The main frame of Figure 4 shows $\left\langle n\right\rangle $ versus
$\lambda_{L}$. Strinkingly, this curve is not monotonic. When $\lambda_{L}$ is
too close to $\lambda_{R}$, the average number of photons in the cavity
collapses, because the element $\left\langle T_{_{-}}\right\vert
h_{so}\left\vert V_{n}\right\rangle $ becomes too small. Two lasing thresholds
appear, one for $\lambda_{L}<\lambda_{R}$ and one for $\lambda_{L}>\lambda
_{R}$. This non-monotonic behavior is directly due to the presence of the
minus signs in Eq.(\ref{alpha1}) and it therefore represents a smoking gun for
coherent Cooper pair injection inside the device. Note that for a typical CPS,
the range of cavity frequency $\nu_{0}$ leading to the lasing effect is rather
broad, because the parameter $\delta$ can be tuned to $\delta_{l}(\nu_{0})$
with the dots' gate voltages. The inset of figure 4 shows the maximum number
of photons in the cavity, obtained at $\delta=\delta_{l}(\nu_{0})$, as a
function of $\nu_{0}$ and $\left\vert \lambda_{L}-\lambda_{R}\right\vert $.
For the parameters of Fig. 3 and in particular $\left\vert \lambda_{L}%
-\lambda_{R}\right\vert =2.5$ $10^{-6}\Delta_{r}$, the range $3.3~GHz<\nu
_{0}<5.9~GHz$ allows one to have $\left\langle n\right\rangle \geq40$.
Interestingly, for $\delta_{r}>2\Delta_{r}$, the state $\left\vert
V_{2}\right\rangle $ is more probable than $\left\vert V_{1}\right\rangle $,
but it cannot produce any lasing effect. In this case, one can imagine to test
the existence of the minus sign in Eq.(\ref{alpha1}) by applying to the CPS a
classical AC gate voltage with frequency $(E_{triplet}-E_{2})/2\pi\hbar$
instead of coupling it to the electric field of a cavity. The current through
the CPS should show a non-monotonic dependence with respect to $\lambda_{L}$
or $\lambda_{R}$. However, the current variations corresponding to this effect
might be very small. The lasing effect discussed in this letter presents the
advantage of providing an intrinsic and powerful amplification mechanism for
the coefficient $\left\langle T_{_{-}}\right\vert h_{so}\left\vert
V_{1}\right\rangle $. This constrasts with a previous proposal where a cavity
was not used \cite{Cerletti}.

Before concluding, we discuss our findings on a more general level. In
principle, spin-flip processes induced by the magnetic coupling $g_{L(R)}$
between the cavity photons and spins in dot $L(R)$ can also lead to the
subradiance effect. Using a generic single orbital model for each dot, which
leads to CPS doubly occupied states $\left\vert \sigma,\sigma^{\prime
}\right\rangle $, we find transition elements between the singlet state
$\left\vert \downarrow,\uparrow\right\rangle -\left\vert \uparrow
,\downarrow\right\rangle $ and the triplet states $\left\vert \uparrow
,\uparrow\right\rangle $ and $\left\vert \downarrow,\downarrow\right\rangle $
which are directly proportionnal to $g_{L}-g_{R}$. This suggests that the
subradiance phenomenon studied here is a property of the injected states
rather than of the spin/photon coupling mechanism. Nevertheless, in practice,
$g_{L(R)}$ is expected to be extremely small \cite{imamoglu}. One thus has to
consider a spin/photon coupling term mediated by the cavity electric field,
like e.g. the term of Eq. (\ref{hso}) caused by spin-orbit coupling. In
principle, our findings can be generalized to other types of quantum dots with
spin-orbit coupling like InAs quantum dots (see e.g. Ref. \cite{Eto}), but the
detailed analysis of these cases goes beyond the scope of the present work. In
our case, we have found a coupling element (\ref{alpha1}) which vanishes for
$\Delta_{K\leftrightarrow K^{\prime}}=0$. However, $\Delta_{K\leftrightarrow
K^{\prime}}\neq0$ is not a fundamental constraint to obtain a subradiant
lasing transition. Indeed, at another working point $\delta\sim-2\Delta_{r}$,
we find a third hybridized even state $\left\vert V_{3}\right\rangle
=\sqrt{1-v_{3}^{2}}\left\vert 0,0\right\rangle +v_{3}\left\vert \mathcal{S}%
\right\rangle $ which can decay radiatively to a triplet state $\left\vert
T_{a}\right\rangle $. One can check that the coupling element $\left\langle
T_{a}\right\vert h_{so}\left\vert V_{3}\right\rangle $ has a subradiant form
but remains finite for $\Delta_{K\leftrightarrow K^{\prime}}=0$. Here, we have
chosen to discuss the lasing transition $\left\vert V_{1}\right\rangle
\rightarrow\left\vert T_{-}\right\rangle $ at $\delta=\delta_{l}(\nu_{0}%
)\sim2\Delta_{r}$ because its frequency is more likely to match with the
cavity frequency in practice (see SM, section D for details).

In conclusion we have shown that the coherence of entangled states produced by
a Cooper pair splitter can be proven by the lasing properties of the device
when coupled to a microwave cavity. The idea to probe the coherence of
electronic states using the tools of cavity QED could be generalized to many
other nanoscale circuits.

\textit{We acknowledge fruitful discussions with J. M. Raimond, J. Klinovaja
and B. Braunecker. This work was financed by the EU-FP7 project
SE2ND[271554].}

\newpage

\begin{center}
{\LARGE SUPPLEMENTAL\ MATERIAL}\ \bigskip
\end{center}

\subsection{A. Hamiltonian of the CPS}

Inside the left and right dots $i\in\{L,R\}$, an electron with spin $\sigma
\in\{\uparrow,\downarrow\}$ can be in the orbital $\tau\in\{K,K^{\prime}\}$,
which is reminiscent from the $K/K^{\prime}$ degeneracy of graphene. We use a
double dot effective hamiltonian which includes the superconducting proximity
effect due to the superconducting contact, i.e.%

\begin{align}
H_{ddot}^{eff}  &  =%
{\displaystyle\sum\limits_{i,\tau,\sigma}}
(\varepsilon+\Delta_{so}\tau\sigma)n_{i\tau\sigma}+H_{int}+\Delta
_{K\leftrightarrow K^{\prime}}%
{\displaystyle\sum\limits_{i,\sigma}}
(d_{iK\sigma}^{\dag}d_{iK^{\prime}\sigma}+d_{iK^{\prime}\sigma}^{\dag
}d_{iK\sigma})\label{H}\\
&  +t_{ee}%
{\displaystyle\sum\limits_{\tau,\sigma}}
(d_{L\tau\sigma}^{\dag}d_{R\tau\sigma}+d_{R\tau\sigma}^{\dag}d_{L\tau\sigma
})+t_{eh}%
{\displaystyle\sum\limits_{\tau}}
\left\{  \left(  d_{L\tau\uparrow}^{\dag}d_{R\overline{\tau}\downarrow}^{\dag
}-d_{L\overline{\tau}\downarrow}^{\dag}d_{R\tau\uparrow}^{\dag}\right)
+h.c.\right\} \nonumber
\end{align}
with $\overline{K}=K^{\prime}$ and $\overline{K^{\prime}}=K$. The term in
$t_{eh}$ accounts for coherent injection of singlet Cooper pairs inside the
double dot \cite{Eldridge}. Taking into account the superconducting contacts
with the term in $t_{eh}$ is valid provided quasiparticle transport between
the superconducting contact and the double dot can be disregarded (see part C
of the Supplemental Material). For simplicity, we assume that the orbital
energies in dots $L$ and $R$ are both equal to $\varepsilon$, which can be
obtained by tuning properly the dots' gate voltages. The term $H_{int}$
accounts for Coulomb charging effects. We assume that there cannot be more
than one electron in each dot, due to a strong intra-dot Coulomb charging
energy. The constant $\Delta_{so}$ corresponds to an effective spin-orbit
coupling \cite{Jespersen}. The term $\Delta_{K\leftrightarrow K^{\prime}}$
describes a coupling between the $K$ and $K^{\prime}$ orbitals of dot $i$, due
to disorder at the level of the carbon nanotube atomic structure
\cite{Liang,Kuemmeth,Jespersen}. The hamiltonian $H_{ddot}^{eff}$ must be
supplemented by the normal leads hamiltonian $H_{leads}=%
{\textstyle\sum\nolimits_{k_{\tau},i,\sigma}}
\varepsilon_{ik_{\tau}}c_{ik_{\tau}\sigma}^{\dag}c_{ik_{\tau}\sigma}+h.c.$ and
the tunnel coupling between the dots and normal leads $H_{t}=%
{\textstyle\sum\nolimits_{k_{\tau},\tau,i,\sigma}}
tc_{ik_{\tau}\sigma}^{\dag}d_{i\tau\sigma}+h.c.$, with $c_{ik_{\tau}\sigma}$
the annihilation operator for an electron with spin $\sigma$ in orbital
$k_{\tau}$ of the normal lead $i\in\{L,R\}$.

\section{B. Expression of the CPS eigenstates}

We now discuss the eigenstates and eigenvectors of hamiltonian (\ref{H}) in
the general case. The diagonalization of hamiltonian (\ref{H}) can be
performed by block in five different subpaces, the two subspaces of states
occupied with a single spin $\sigma\in\{\uparrow,\downarrow\}$, the two
subspaces of states occupied with two equal spins $\sigma\in\{\uparrow
,\downarrow\}$, and the subspace comprising the empty state and states
occupied with two opposite spins.

\subsection{B.1. Subspace of singly occupied states with spin $\sigma$}

One can treat separately the subspace of singly occupied states with spin
$\uparrow$ and the subspace of singly occupied states with spin $\downarrow$.
For a given spin direction $\sigma\in\{\uparrow,\downarrow\}$, the
eigenenergies and corresponding eigenvectors of $H_{ddot}^{eff}$ are:%
\[%
\begin{tabular}
[c]{|l|l|}\hline
eigenenergy & eigenvector\\\hline
$\varepsilon_{1\sigma}=\varepsilon-t_{ee}-\Delta_{r}$ & $\left\vert
s_{1\sigma}\right\rangle =\frac{1}{2}\sqrt{1-\sigma\frac{\Delta_{so}}%
{\Delta_{r}}}\left(  \left\vert K\sigma,0\right\rangle -\left\vert
0,K\sigma\right\rangle \right)  +\frac{\Delta_{K/K^{\prime}}}{2\Delta_{r}%
\sqrt{1-\sigma\frac{\Delta_{so}}{\Delta_{r}}}}\left(  \left\vert 0,K^{\prime
}\sigma\right\rangle -\left\vert K^{\prime}\sigma,0\right\rangle \right)
$\\\hline
$\varepsilon_{2\sigma}=\varepsilon+t_{ee}-\Delta_{r}$ & $\left\vert
s_{2\sigma}\right\rangle =-\frac{1}{2}\sqrt{1-\sigma\frac{\Delta_{so}}%
{\Delta_{r}}}\left(  \left\vert K\sigma,0\right\rangle +\left\vert
0,K\sigma\right\rangle \right)  +\frac{\Delta_{K/K^{\prime}}}{2\Delta_{r}%
\sqrt{1-\sigma\frac{\Delta_{so}}{\Delta_{r}}}}\left(  \left\vert 0,K^{\prime
}\sigma\right\rangle +\left\vert K^{\prime}\sigma,0\right\rangle \right)
$\\\hline
$\varepsilon_{3\sigma}=\varepsilon-t_{ee}+\Delta_{r}$ & $\left\vert
s_{3\sigma}\right\rangle =-\frac{1}{2}\sqrt{1+\sigma\frac{\Delta_{so}}%
{\Delta_{r}}}\left(  \left\vert K\sigma,0\right\rangle -\left\vert
0,K\sigma\right\rangle \right)  +\frac{\Delta_{K/K^{\prime}}}{2\Delta_{r}%
\sqrt{1+\sigma\frac{\Delta_{so}}{\Delta_{r}}}}\left(  \left\vert 0,K^{\prime
}\sigma\right\rangle \}-\left\vert K^{\prime}\sigma,0\right\rangle \right)
$\\\hline
$\varepsilon_{4\sigma}=\varepsilon+t_{ee}+\Delta_{r}$ & $\left\vert
s_{4\sigma}\right\rangle =\frac{1}{2}\sqrt{1+\sigma\frac{\Delta_{so}}%
{\Delta_{r}}}\left(  \left\vert K\sigma,0\right\rangle +\left\vert
0,K\sigma\right\rangle \right)  +\frac{\Delta_{K/K^{\prime}}}{2\Delta_{r}%
\sqrt{1+\sigma\frac{\Delta_{so}}{\Delta_{r}}}}\left(  \left\vert 0,K^{\prime
}\sigma\right\rangle \}+\left\vert K^{\prime}\sigma,0\right\rangle \right)
$\\\hline
\end{tabular}
\ \ \
\]
with $\Delta_{r}=\sqrt{\Delta_{so}^{2}+\Delta_{K/K^{\prime}}^{2}}$. Note that
$t_{ee}$ occurs in the expressions of the eigenenergies but not in the
expressions of the eigenvectors because we have assumed that the left and
rights dot have the same orbital energy $\varepsilon$.

\subsection{B.2. Subspace of states occupied with two equal spins $\sigma$}

One can treat separately the subspace of states occupied with two equal spins
$\uparrow$ and the subspace of states occupied with two equal spins
$\downarrow$. For a given spin direction $\sigma\in\{\uparrow,\downarrow\}$,
the eigenenergies and corresponding eigenvectors of $H_{ddot}^{eff}$ are:
\[%
\begin{tabular}
[c]{|l|l|}\hline
eigenenergy & eigenvector\\\hline
$\delta$ & $\left\vert \tilde{S}_{1\sigma}\right\rangle =\frac{\Delta
_{K/K^{\prime}}}{2\tilde{\Delta}_{r}}\left(  \left\vert K^{\prime}%
\sigma,K^{\prime}\sigma\right\rangle -\left\vert K\sigma,K\sigma\right\rangle
\right)  +\sigma\frac{\Delta_{so}}{\tilde{\Delta}_{r}}\left\vert K^{\prime
}\sigma,K\sigma\right\rangle $\\\hline
$\delta$ & $\left\vert \tilde{S}_{2\sigma}\right\rangle =\frac{\Delta
_{K/K^{\prime}}}{2\tilde{\Delta}_{r}}\left(  \left\vert K^{\prime}%
\sigma,K^{\prime}\sigma\right\rangle -\left\vert K\sigma,K\sigma\right\rangle
\right)  +\sigma\frac{\Delta_{so}}{\tilde{\Delta}_{r}}\left\vert
K\sigma,K^{\prime}\sigma\right\rangle $\\\hline
$\delta-2\Delta_{r}$ & $\left\vert \tilde{S}_{3\sigma}\right\rangle =\frac
{1}{2}\left(  \sigma\frac{\Delta_{so}}{\Delta_{r}}-1\right)  \left\vert
K\sigma,K\sigma\right\rangle -\frac{1}{2}\left(  1+\sigma\frac{\Delta_{so}%
}{\Delta_{r}}\right)  \left\vert K^{\prime}\sigma,K^{\prime}\sigma
\right\rangle +\frac{\Delta_{K/K^{\prime}}}{2\Delta_{r}}\left(  \left\vert
K\sigma,K^{\prime}\sigma\right\rangle +\left\vert K^{\prime}\sigma
,K\sigma\right\rangle \right)  $\\\hline
$\delta+2\Delta_{r}$ & $\left\vert \tilde{S}_{4\sigma}\right\rangle =\frac
{1}{2}\left(  1+\sigma\frac{\Delta_{so}}{\Delta_{r}}\right)  \left\vert
K\sigma,K\sigma\right\rangle +\frac{1}{2}\left(  1-\sigma\frac{\Delta_{so}%
}{\Delta_{r}}\right)  \left\vert K^{\prime}\sigma,K^{\prime}\sigma
\right\rangle +\frac{\Delta_{K/K^{\prime}}}{2\Delta_{r}}\left(  \left\vert
K\sigma,K^{\prime}\sigma\right\rangle +\left\vert K^{\prime}\sigma
,K\sigma\right\rangle \right)  $\\\hline
\end{tabular}
\
\]

with $\tilde{\Delta}_{r}=\sqrt{\Delta_{so}^{2}+(\Delta_{K/K^{\prime}}^{2}/2)}%
$. These states correspond to generalized triplet states with spin 1.

\subsection{B.3. Subspace of states occupied with two opposite spins}

We now discuss the subspace of states comprising the empty state $\left\vert
0,0\right\rangle $ and states occupied with two opposite spins. It is
practical to first define the eigenenergies and eigenvectors of $H_{ddot}%
^{eff}$ for $t_{eh}=0$:%
\[%
\begin{tabular}
[c]{|l|l|}\hline
eigenenergy & eigenvector\\\hline
$0$ & $\left\vert 0,0\right\rangle $\\\hline
$\delta$ & $\left\vert s_{1}\right\rangle =\frac{\Delta_{K/K^{\prime}}%
}{2\tilde{\Delta}_{r}}\left(  \left\vert \mathcal{C}_{-}(K\downarrow
,K^{\prime}\uparrow)\right\rangle -\left\vert \mathcal{C}_{-}(K^{\prime
}\downarrow,K\uparrow)\right\rangle \right)  +\frac{\Delta_{so}}{\tilde
{\Delta}_{r}}\left\vert \mathcal{C}_{-}(K^{\prime}\downarrow,K^{\prime
}\uparrow)\right\rangle $\\\hline
$\delta$ & $\left\vert t_{1}\right\rangle =\frac{\Delta_{K/K^{\prime}}%
}{2\tilde{\Delta}_{r}}\left(  \left\vert \mathcal{C}_{+}(K\downarrow
,K^{\prime}\uparrow)\right\rangle -\left\vert \mathcal{C}_{+}(K^{\prime
}\downarrow,K\uparrow)\right\rangle \right)  +\frac{\Delta_{so}}{\tilde
{\Delta}_{r}}\left\vert \mathcal{C}_{+}(K^{\prime}\downarrow,K^{\prime
}\uparrow)\right\rangle $\\\hline
$\delta$ & $\left\vert s_{2}\right\rangle =\frac{\Delta_{K/K^{\prime}}%
}{2\tilde{\Delta}_{r}}\left(  \left\vert \mathcal{C}_{-}(K\downarrow
,K^{\prime}\uparrow)\right\rangle -\left\vert \mathcal{C}_{-}(K^{\prime
}\downarrow,K\uparrow)\right\rangle \right)  +\frac{\Delta_{so}}{\tilde
{\Delta}_{r}}\left\vert \mathcal{C}_{-}(K\downarrow,K\uparrow)\right\rangle
$\\\hline
$\delta$ & $\left\vert t_{2}\right\rangle =\frac{\Delta_{K/K^{\prime}}%
}{2\tilde{\Delta}_{r}}\left(  \left\vert \mathcal{C}_{+}(K\downarrow
,K^{\prime}\uparrow)\right\rangle -\left\vert \mathcal{C}_{+}(K^{\prime
}\downarrow,K\uparrow)\right\rangle \right)  +\frac{\Delta_{so}}{\tilde
{\Delta}_{r}}\left\vert \mathcal{C}_{+}(K\downarrow,K\uparrow)\right\rangle
$\\\hline
$\delta-2\Delta_{r}$ & $2\left\vert s_{3}\right\rangle =(\frac{\Delta_{so}%
}{\Delta_{r}}-1)\left\vert \mathcal{C}_{-}(K\uparrow,K^{\prime}\downarrow
)\right\rangle -(\frac{\Delta_{so}}{\Delta_{r}}+1)\left\vert \mathcal{C}%
_{-}(K^{\prime}\uparrow,K\downarrow)\right\rangle +\frac{\Delta_{K/K^{\prime}%
}}{\Delta_{r}}\left(  \left\vert \mathcal{C}_{-}(K\uparrow,K\downarrow
)\right\rangle +\left\vert \mathcal{C}_{-}(K^{\prime}\uparrow,K^{\prime
}\downarrow)\right\rangle \right)  $\\\hline
$\delta-2\Delta_{r}$ & $2\left\vert t_{3}\right\rangle =(\frac{\Delta_{so}%
}{\Delta_{r}}-1)\left\vert \mathcal{C}_{+}(K\uparrow,K^{\prime}\downarrow
)\right\rangle -(\frac{\Delta_{so}}{\Delta_{r}}+1)\left\vert \mathcal{C}%
_{+}(K^{\prime}\uparrow,K\downarrow)\right\rangle +\frac{\Delta_{K/K^{\prime}%
}}{\Delta_{r}}\left(  \left\vert \mathcal{C}_{+}(K\uparrow,K\downarrow
)\right\rangle +\left\vert \mathcal{C}_{+}(K^{\prime}\uparrow,K^{\prime
}\downarrow)\right\rangle \right)  $\\\hline
$\delta+2\Delta_{r}$ & $2\left\vert s_{4}\right\rangle =(\frac{\Delta_{so}%
}{\Delta_{r}}+1)\left\vert \mathcal{C}_{-}(K\uparrow,K^{\prime}\downarrow
)\right\rangle +(1-\frac{\Delta_{so}}{\Delta_{r}})\left\vert \mathcal{C}%
_{-}(K^{\prime}\uparrow,K\downarrow)\right\rangle +\frac{\Delta_{K/K^{\prime}%
}}{\Delta_{r}}\left(  \left\vert \mathcal{C}_{-}(K\uparrow,K\downarrow
)\right\rangle +\left\vert \mathcal{C}_{-}(K^{\prime}\uparrow,K^{\prime
}\downarrow)\right\rangle \right)  $\\\hline
$\delta+2\Delta_{r}$ & $2\left\vert t_{4}\right\rangle =(\frac{\Delta_{so}%
}{\Delta_{r}}+1)\left\vert \mathcal{C}_{+}(K\uparrow,K^{\prime}\downarrow
)\right\rangle +(1-\frac{\Delta_{so}}{\Delta_{r}})\left\vert \mathcal{C}%
_{+}(K^{\prime}\uparrow,K\downarrow)\right\rangle +\frac{\Delta_{K/K^{\prime}%
}}{\Delta_{r}}\left(  \left\vert \mathcal{C}_{+}(K\uparrow,K\downarrow
)\right\rangle +\left\vert \mathcal{C}_{+}(K^{\prime}\uparrow,K^{\prime
}\downarrow)\right\rangle \right)  $\\\hline
\end{tabular}
\ \ \ \ \ \
\]
where $\left\vert \mathcal{C}_{\pm}(\tau\sigma,\tau^{\prime}\sigma^{\prime
})\right\rangle =(\left\vert \tau\sigma,\tau^{\prime}\sigma^{\prime
}\right\rangle \pm\left\vert \tau^{\prime}\sigma^{\prime},\tau\sigma
\right\rangle )/\sqrt{2}$. From the definition of $\left\vert \mathcal{C}%
_{\pm}(\tau\sigma,\tau^{\prime}\sigma^{\prime})\right\rangle $, one can view
the states $\left\vert s_{1}\right\rangle $, $\left\vert s_{2}\right\rangle $,
$\left\vert s_{3}\right\rangle $ and $\left\vert s_{4}\right\rangle $ as
generalized singlet states whereas $\left\vert t_{1}\right\rangle $,
$\left\vert t_{2}\right\rangle $, $\left\vert t_{3}\right\rangle $ and
$\left\vert t_{4}\right\rangle $ can be viewed as generalized triplet states
with total spin 0. In the subspace \{$\left\vert 0,0\right\rangle
$,$\left\vert s_{1}\right\rangle ,\left\vert t_{1}\right\rangle ,\left\vert
s_{2}\right\rangle ,\left\vert t_{2}\right\rangle ,\left\vert s_{3}%
\right\rangle ,\left\vert t_{3}\right\rangle ,\left\vert s_{4}\right\rangle
,\left\vert t_{4}\right\rangle $\}, the hamiltonian (\ref{H}) writes exactly:%
\begin{align*}
\tilde{H}_{ddot}^{eff}  &  =\delta(\left\vert s_{1}\right\rangle \left\langle
s_{1}\right\vert +\left\vert s_{2}\right\rangle \left\langle s_{2}\right\vert
+\left\vert t_{1}\right\rangle \left\langle t_{1}\right\vert +\left\vert
t_{2}\right\rangle \left\langle t_{2}\right\vert )+(\delta-2\Delta_{r})\left(
\left\vert t_{3}\right\rangle \left\langle t_{3}\right\vert +\left\vert
s_{3}\right\rangle \left\langle s_{3}\right\vert \right)  +(\delta+2\Delta
_{r})\left(  \left\vert t_{4}\right\rangle \left\langle t_{4}\right\vert
+\left\vert s_{4}\right\rangle \left\langle s_{4}\right\vert \right) \\
&  +\sqrt{2}t_{eh}\left(  \left\vert s_{4}\right\rangle \left\langle
0,0\right\vert +\left\vert 0,0\right\rangle \left\langle s_{4}\right\vert
-\left\vert s_{3}\right\rangle \left\langle 0,0\right\vert -\left\vert
0,0\right\rangle \left\langle s_{3}\right\vert \right)
\end{align*}
Thus, the states $\left\vert s_{1}\right\rangle $, $\left\vert s_{2}%
\right\rangle $, $\left\vert t_{1}\right\rangle $,$\left\vert t_{2}%
\right\rangle $, $\left\vert t_{3}\right\rangle $ and $\left\vert
t_{4}\right\rangle $ are eigenstates of $H_{ddot}^{eff}$ while the states
$\left\vert s_{3}\right\rangle $, and $\left\vert s_{4}\right\rangle $ and are
hybridized with $\left\vert 0,0\right\rangle $. This hybridization leads to
three eigenstates $\left\vert V_{1}\right\rangle $, $\left\vert V_{2}%
\right\rangle $ and $\left\vert V_{3}\right\rangle $ with energy $E_{1}$,
$E_{2}$ and $E_{3}$. Figure 1 shows the energy of the different doubly
occupied eigenstates (with total spin 1 or 0, i.e. given by sections B.2 or
B.3) as a function of $\delta$. The triplet states have energies $\delta$,
$\delta-2\Delta_{r}$ or $\delta+2\Delta_{r}$ (blue lines), while the energies
$E_{1}$, $E_{2}$ and $E_{3}$ (red, green and pink lines) show two
anticrossings at $\delta=-2\Delta_{r}$ and $\delta=2\Delta_{r}$.

\begin{center}
\begin{figure}[h]
\includegraphics[width=0.5\linewidth]{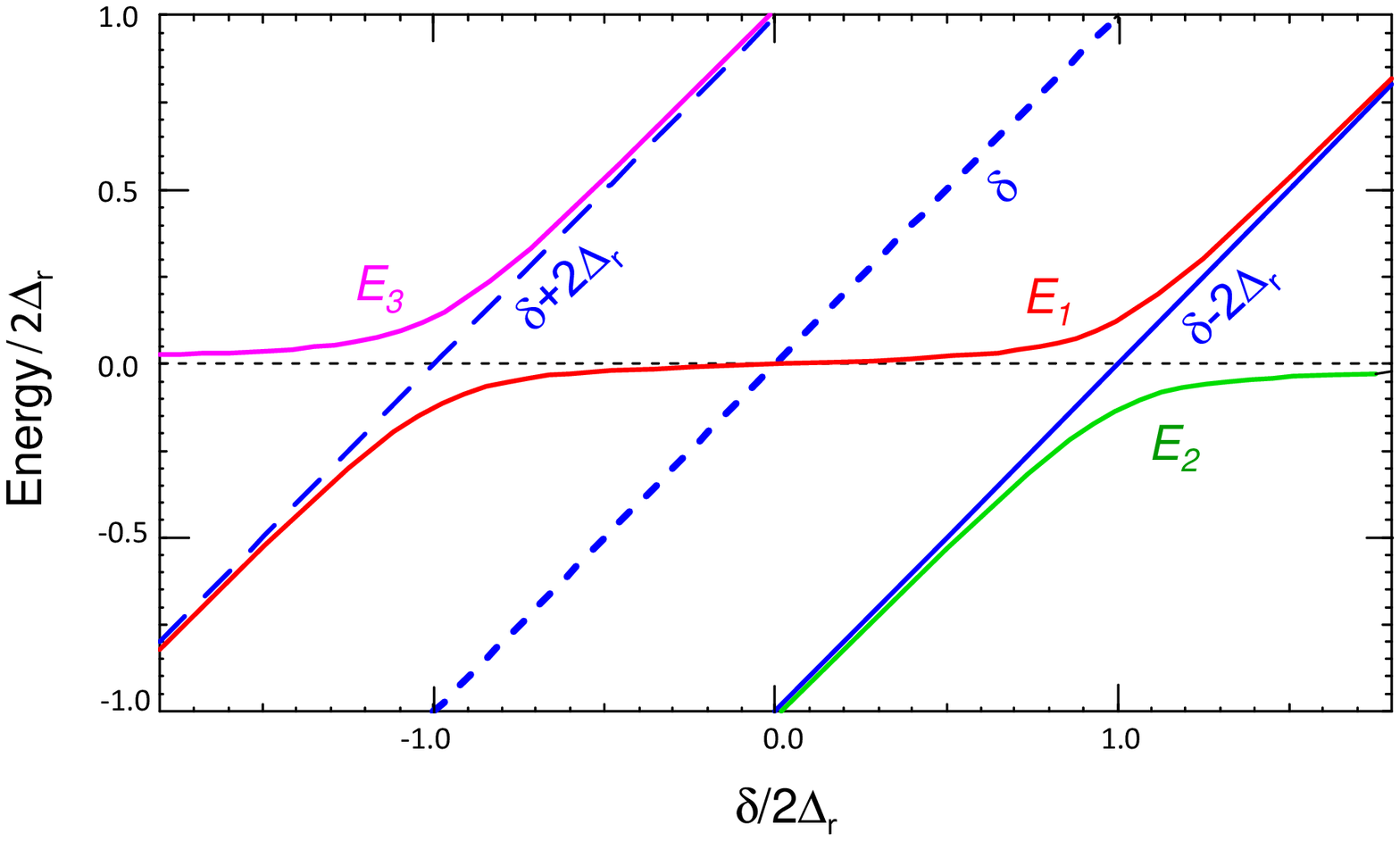}\caption{Energy of the
different doubly occupied eigenstates (with total spin 0 or 1) as a function
of $\delta$.}%
\end{figure}
\end{center}

\subsection{B.4. Expression of the CPS doubly occupied eigenstates for
$\delta\sim2\Delta_{r}$}

In the main text, we work near the right anticrossing in Figure 1 ($\delta
\sim2\Delta_{r}$). For $\delta\sim2\Delta_{r}$, a simplified expression of the
relevant doubly occupied eigenstates of $H_{ddot}^{eff}$ can be obtained by
performing a diagonalization in the subpace\ \{$\left\vert 0,0\right\rangle $,
$\left\vert s_{3}\right\rangle $,$\left\vert t_{3}\right\rangle $, $\left\vert
\tilde{S}_{3\uparrow}\right\rangle $,$\left\vert \tilde{S}_{3\downarrow
}\right\rangle $\}. This leads to the CPS eigenstates $\left\vert
V_{1}\right\rangle $, $\left\vert V_{2}\right\rangle $, $\left\vert
T_{0}\right\rangle $, $\left\vert T_{+}\right\rangle $ and $\left\vert
T_{-}\right\rangle $ discussed in the main text, which are defined by:%
\[%
\begin{tabular}
[c]{|l|l|}\hline
eigenenergy & eigenvector\\\hline
$\delta-2\Delta_{r}$ & $\left\vert T_{0}\right\rangle =\left\vert
t_{3}\right\rangle =\sum_{\sigma}\left\{  \frac{1}{2}(\sigma\frac{\Delta_{so}%
}{\Delta_{r}}-1)\left\vert \mathcal{C}_{+}(K\sigma,K^{\prime}\bar{\sigma
})\right\rangle \right\}  +\frac{\Delta_{K/K^{\prime}}}{2\Delta_{r}}%
{\displaystyle\sum\limits_{\tau}}
\left\vert \mathcal{C}_{+}(\tau\uparrow,\tau\downarrow)\right\rangle $\\\hline
$\delta-2\Delta_{r}$ & $\left\vert T_{+}\right\rangle =\frac{\left\vert
\tilde{S}_{3\uparrow}\right\rangle -\left\vert \tilde{S}_{3\downarrow
}\right\rangle }{\sqrt{2}}=\sum_{\sigma}\left\{  \frac{1}{2}\left(
\frac{\Delta_{so}}{\Delta_{r}}-\sigma\right)  \frac{\left\vert K\sigma
,K\sigma\right\rangle -\left\vert K^{\prime}\bar{\sigma},K^{\prime}\bar
{\sigma}\right\rangle }{\sqrt{2}}+\sigma\frac{\Delta_{K/K^{\prime}}}%
{2\Delta_{r}}\left\vert \mathcal{C}_{+}(K\sigma,K^{\prime}\sigma)\right\rangle
\right\}  $\\\hline
$\delta-2\Delta_{r}$ & $\left\vert T_{-}\right\rangle =\frac{\left\vert
\tilde{S}_{3\uparrow}\right\rangle +\left\vert \tilde{S}_{3\downarrow
}\right\rangle }{\sqrt{2}}=\sum_{\sigma}\left\{  \frac{1}{2}\left(
\frac{\Delta_{so}}{\Delta_{r}}\sigma-1\right)  \frac{\left\vert K\sigma
,K\sigma\right\rangle +\left\vert K^{\prime}\bar{\sigma},K^{\prime}\bar
{\sigma}\right\rangle }{\sqrt{2}}+\frac{\Delta_{K/K^{\prime}}}{2\Delta_{r}%
}\left\vert \mathcal{C}_{+}(K\sigma,K^{\prime}\sigma)\right\rangle \right\}
$\\\hline
$E_{1}$ & $\left\vert V_{1}\right\rangle =\sqrt{1-v_{1}^{2}}\left\vert
0,0\right\rangle +v_{1}\left\vert \mathcal{S}\right\rangle $\\\hline
$E_{2}$ & $\left\vert V_{2}\right\rangle =\sqrt{1-v_{2}^{2}}\left\vert
0,0\right\rangle +v_{2}\left\vert \mathcal{S}\right\rangle $\\\hline
\end{tabular}
\ \ \ \ \ \
\]
with%
\[
E_{i}=\frac{1}{2}\left(  \delta-2\Delta_{r}-(-1)^{n}\sqrt{8t_{eh}^{2}%
+(\delta-2\Delta_{r})^{2}}\right)
\]%
\[
v_{n}=\frac{2t_{eh}}{\sqrt{8t_{eh}^{2}+(\delta-2\Delta_{r})(\delta-2\Delta
_{r}+(-1)^{n}\sqrt{8t_{eh}^{2}+(\delta-2\Delta_{r})^{2}})}}%
\]
and%
\[
\left\vert \mathcal{S}\right\rangle =\left\vert s_{3}\right\rangle
=\sum_{\sigma}\left\{  \frac{1}{2}(\frac{\Delta_{so}}{\Delta_{r}}%
-\sigma)\left\vert \mathcal{C}_{-}(K\sigma,K^{\prime}\bar{\sigma
})\right\rangle \right\}  +\frac{\Delta_{K/K^{\prime}}}{2\Delta_{r}}%
{\displaystyle\sum\limits_{\tau}}
\left\vert \mathcal{C}_{-}(\tau\uparrow,\tau\downarrow)\right\rangle
\]
Note that $\sigma=\pm1$ stands for $\sigma\in\{\uparrow,\downarrow\}$ in
algebraic expressions. The state $\left\vert T_{0}\right\rangle $ corresponds
to a generalized triplet state with zero spin, $\left\vert T_{-}\right\rangle
$ and $\left\vert T_{+}\right\rangle $ correspond to a coherent mixture of two
triplet states with finite spin, and the state $\left\vert \mathcal{S}%
\right\rangle $ corresponds to a generalized spin-singlet state. Note that
without disorder ($\Delta_{K\leftrightarrow K^{\prime}}=0$), $\left\vert
\mathcal{S}\right\rangle $ has components in $\left\vert \mathcal{C}_{+}%
(\tau\downarrow,\overline{\tau}\uparrow)\right\rangle $ only and $\left\vert
T_{-}\right\rangle $ has components in $\left\vert \tau\sigma,\tau
\sigma\right\rangle $ only. In the presence of disorder, $\left\vert
\mathcal{S}\right\rangle $ also includes components in $\mathcal{C}%
_{-}(\left\vert \tau\uparrow,\tau\downarrow\right\rangle )$, and $\left\vert
T_{-}\right\rangle $ has also components in $\left\vert K\sigma,K^{\prime
}\sigma\right\rangle $ and $\left\vert K^{\prime}\sigma,K\sigma\right\rangle
$. This enables a coupling between the states $\left\vert V_{1(2)}%
\right\rangle $\ and $\left\vert T_{-}\right\rangle $ through $H_{so}$.

\section{C. Bias voltage window}

The hamiltonian $H_{ddot}^{eff}$ can be used provided there is no
quasiparticle transport between the superconducting lead and the dots. This
requires
\begin{equation}
-\Delta<-eV<\Delta\label{aa}%
\end{equation}
In the main text, we furthermore assume that electrons can go from the double
dot to the normal metal leads but not the reverse. This is true provided the
bias voltage $V$ belongs to a certain range which we derive below.

\textbf{(1)} We recall that the double dot singly occupied states $\left\vert
s_{i\sigma}\right\rangle $ have energies $\varepsilon_{i\sigma}$, with
$i\in\{1,2,3,4\}$, given in section B.1 of the supplemental material. We
assume $\delta\sim2\Delta_{r}$, so that the states $\left\vert V_{1}%
\right\rangle $ and $\left\vert V_{2}\right\rangle $ have significant
components in $\left\vert 0,0\right\rangle $ and $\left\vert \mathcal{S}%
\right\rangle $ and their energies $E_{1}$ and $E_{2}$ are well approximated
by the expressions given in section B.4. Since $\left\vert V_{1}\right\rangle
$ and $\left\vert V_{2}\right\rangle $ have components in $\left\vert
\mathcal{S}\right\rangle $, they can decay towards the singly occupied state
$\left\vert s_{i\sigma}\right\rangle $ while an electron is transferred to the
normal metal leads (and the reverse process is forbidden) if
\begin{equation}
-eV<E_{1(2)}-\varepsilon_{i\sigma}-\alpha k_{B}T \label{a}%
\end{equation}
for $i\in\{1,2,3,4\}$. Above, $\alpha$ is a dimensionless factor of order 1
which takes into account the temperature broadening of the levels.

\textbf{(2)} Since $\left\vert V_{1}\right\rangle $ and$\left\vert
V_{2}\right\rangle $ have also components in $\left\vert 0,0\right\rangle $, a
singly occupied state can decay towards $\left\vert V_{1}\right\rangle $ or
$\left\vert V_{2}\right\rangle $ while an electron is transferred to the
normal leads (and the reverse process is forbidden) if
\begin{equation}
-eV<\varepsilon_{i\sigma}-E_{1(2)}-\alpha k_{B}T \label{b}%
\end{equation}
for $i\in\{1,2,3,4\}$.

\textbf{(3) }Since we assume $\delta\sim2\Delta_{r}$, the state $\left\vert
V_{3}\right\rangle $ has a negligible component in $\left\vert
0,0\right\rangle $. Hence, it can be considered as a pure doubly occupied
state, with energy $E_{3}\simeq\delta+2\Delta$. The other doubly occupied
states have energies $\delta-2\Delta_{r}$, $\delta+2\Delta_{r}$, or $\delta$
(see section B.2 and B.3). The doubly occupied states (including $\left\vert
V_{3}\right\rangle $) can relax to the singly occupied state $\left\vert
s_{i\sigma}\right\rangle $ while an electron is transferred to the normal
leads (and the reverse processes are forbidden) if
\begin{equation}
-eV<\delta\pm2\Delta_{r}-\varepsilon_{i\sigma}-\alpha k_{B}T \label{c}%
\end{equation}
and
\begin{equation}
-eV<\delta-\varepsilon_{i\sigma}-\alpha k_{B}T \label{d}%
\end{equation}
Since $E_{2}<\delta-2\Delta_{r}<E_{1},\delta<\delta+2\Delta_{r}<E_{3}$ and
$\varepsilon_{1\sigma}<\varepsilon_{2\sigma},\varepsilon_{3\sigma}%
<\varepsilon_{4\sigma}$, the combination of Eqs. (\ref{aa}), (\ref{a}),
(\ref{b}), (\ref{c}) and (\ref{d}) yields the constraint%
\[
-\Delta<-eV<\min(E_{2}-\varepsilon_{4\sigma},\varepsilon_{1\sigma}%
-E_{1})-\alpha k_{B}T
\]
We note $\delta=2\varepsilon$. For $\delta\sim2\Delta_{r}$, one has%
\[
E_{2}-\varepsilon_{4\sigma}=-2\Delta_{r}-t_{ee}-\frac{1}{2}\sqrt{8t_{eh}%
^{2}+(\delta-2\Delta_{r})^{2}}=\varepsilon_{1\sigma}-E_{1}-2\Delta_{r}%
\]
We conclude that we have to satisfy%
\[
-\Delta<-eV<-2\Delta_{r}-t_{ee}-\frac{1}{2}\sqrt{8t_{eh}^{2}+(\delta
-2\Delta_{r})^{2}}-\alpha k_{B}T
\]
with $\Delta$ the BCS gap of the superconducting contact. Since the
temperature $T$ is much smaller than $\Delta_{r}$ in a typical experiment, we
simplify this criterion as%
\[
-\Delta<-eV<-2\Delta_{r}-t_{ee}-\frac{1}{2}\sqrt{8t_{eh}^{2}+(\delta
-2\Delta_{r})^{2}}%
\]
With the parameters of Fig.2, $t_{ee}\ll\Delta_{r}$, $\delta\sim2\Delta_{r}$
and $\Delta$ the BCS gap of NbN, this gives $1.8~\mathrm{meV}\lesssim
eV\lesssim3~\mathrm{meV}$.

To populate the state $\left\vert V_{3}\right\rangle $ and the triplet states
other than $\left\vert T_{-}\right\rangle $, one needs to have a transition
from a singly occupied state to one of these states. In the regime we consider
above, this is not possible since this would require an electron to go from
the normal metal leads to the double dot. Therefore, the state $\left\vert
V_{3}\right\rangle $ and the triplet states other than $\left\vert
T_{-}\right\rangle $ are not active. In contrast, the state $\left\vert
T_{-}\right\rangle $ can be populated due to lasing transitions $\left\vert
V_{1}\right\rangle \rightarrow\left\vert T_{-}\right\rangle $.

\section{D. Effect of the spin-orbit coupling}

In order to discuss the effect of the term $h_{so}$ appearing in Eq. (2) of
the main text, it is practical to redefine the eigenvectors of $H_{ddot}%
^{eff}$ in the subspace of states occupied with two equal spins as
\[%
\begin{tabular}
[c]{|l|l|}\hline
eigenenergy & eigenvectors\\\hline
$\delta$ & $\left\vert T_{a}\right\rangle =\frac{\alpha_{+}}{\sqrt{\alpha
_{-}^{2}+\alpha_{+}^{2}}}\frac{\left\vert \tilde{S}_{1\uparrow}\right\rangle
-\left\vert \tilde{S}_{2\downarrow}\right\rangle }{\sqrt{2}}-\frac{\alpha_{-}%
}{\sqrt{\alpha_{-}^{2}+\alpha_{+}^{2}}}\frac{\left\vert \tilde{S}_{2\uparrow
}\right\rangle -\left\vert \tilde{S}_{1\downarrow}\right\rangle }{\sqrt{2}}%
$\\\hline
$\delta$ & $\left\vert T_{b}\right\rangle =\frac{\alpha_{-}}{\sqrt{\alpha
_{-}^{2}+\alpha_{+}^{2}}}\frac{\left\vert \tilde{S}_{1\uparrow}\right\rangle
-\left\vert \tilde{S}_{2\downarrow}\right\rangle }{\sqrt{2}}-\frac{\alpha_{+}%
}{\sqrt{\alpha_{-}^{2}+\alpha_{+}^{2}}}\frac{\left\vert \tilde{S}_{2\uparrow
}\right\rangle -\left\vert \tilde{S}_{1\downarrow}\right\rangle }{\sqrt{2}}%
$\\\hline
$\delta$ & $\left\vert T_{1-}\right\rangle =\frac{\left\vert \tilde
{S}_{1\uparrow}\right\rangle +\left\vert \tilde{S}_{2\downarrow}\right\rangle
}{\sqrt{2}}$\\\hline
$\delta$ & $\left\vert T_{2-}\right\rangle =\frac{\left\vert \tilde
{S}_{2\uparrow}\right\rangle +\left\vert \tilde{S}_{1\downarrow}\right\rangle
}{\sqrt{2}}$\\\hline
$\delta-2\Delta_{r}$ & $\left\vert T_{+}\right\rangle $ {\small already
defined in section B.4}\\\hline
$\delta-2\Delta_{r}$ & $\left\vert T_{-}\right\rangle $ {\small already
defined in section B.4}\\\hline
$\delta+2\Delta_{r}$ & $\left\vert \tilde{S}_{4\uparrow}\right\rangle
$\\\hline
$\delta+2\Delta_{r}$ & $\left\vert \tilde{S}_{4\downarrow}\right\rangle
$\\\hline
\end{tabular}
\ \ \ \ \ \ \
\]
with%
\[
\alpha_{\mp}=\lambda_{L}-\lambda_{R}\mp\frac{\Delta_{so}}{\Delta_{r}}%
(\lambda_{L}+\lambda_{R})
\]
For $\delta\sim2\Delta_{r}$ and the bias voltage conditions considered in
section C, the states $\left\vert V_{1}\right\rangle $ and $\left\vert
V_{2}\right\rangle $ can be populated but not $\left\vert V_{3}\right\rangle
$. Only two of the triplet states, namely $\left\vert T_{-}\right\rangle $ and
$\left\vert T_{b}\right\rangle $, are coupled to $\left\vert V_{1}%
\right\rangle $ and $\left\vert V_{2}\right\rangle $ by $h_{so}$. One has%
\begin{equation}
\left\langle T_{_{-}}\right\vert h_{so}\left\vert V_{1(2)}\right\rangle
=\mathbf{i}v_{1(2)}\frac{\Delta_{K\leftrightarrow K^{\prime}}}{\Delta_{r}%
}(\lambda_{L}-\lambda_{R}) \label{alpha1b}%
\end{equation}
which corresponds to Eq. (3) of the main text, and
\begin{equation}
\left\langle T_{b}\right\vert h_{so}\left\vert V_{1(2)}\right\rangle
=\frac{\mathbf{i}v_{1(2)}\Delta_{so}^{2}(\lambda_{R}^{2}-\lambda_{L}^{2}%
)}{\sqrt{\tilde{\Delta}_{r}\left(  \Delta_{so}^{2}(\lambda_{L}^{2}+\lambda
_{R}^{2})+\frac{\Delta_{K/K^{\prime}}^{2}}{2}(\lambda_{L}-\lambda_{R}%
)^{2}\right)  }} \label{c2}%
\end{equation}
The couplings to $\left\vert T_{-}\right\rangle $ and $\left\vert
T_{b}\right\rangle $ are both subradiant since they vanish for $\lambda
_{L}=\lambda_{R}$. For $\Delta_{K\leftrightarrow K^{\prime}}=0$, the coupling
between $\left\vert T_{-}\right\rangle $ and $\left\vert V_{1(2)}\right\rangle
$ vanishes, whereas the coupling between $\left\vert T_{b}\right\rangle $ and
$\left\vert V_{1(2)}\right\rangle $ persists:%
\begin{equation}
\lim_{\Delta_{K\leftrightarrow K^{\prime}}\rightarrow0}\left\langle
T_{b}\right\vert h_{so}\left\vert V_{1(2)}\right\rangle =\frac{\mathbf{i}%
v_{1(2)}(\lambda_{R}^{2}-\lambda_{L}^{2})}{\sqrt{\lambda_{L}^{2}+\lambda
_{R}^{2}}} \label{ll}%
\end{equation}
We conclude that it is not necessary to use a finite $\Delta_{K\leftrightarrow
K^{\prime}}$ to obtain a coupling between $\left\vert V_{1(2)}\right\rangle $
and a triplet state. It is nevertheless impossible to use the transitions
$\left\vert V_{1(2)}\right\rangle \leftrightarrow\left\vert T_{b}\right\rangle
$ for lasing since $\left\vert T_{b}\right\rangle $ is higher in energy than
$\left\vert V_{1}\right\rangle $ and $\left\vert V_{2}\right\rangle $.

For $\delta\sim-2\Delta_{r}$, the state $\left\vert V_{3}\right\rangle $ can
be populated. When $\Delta_{K\leftrightarrow K^{\prime}}=0$, we find that only
one of the triplet states, namely $\left\vert T_{a}\right\rangle $, is coupled
to $\left\vert V_{3}\right\rangle $, with a coupling element%
\begin{equation}
\lim_{\Delta_{K\leftrightarrow K^{\prime}}\rightarrow0}\left\langle
T_{a}\right\vert h_{so}\left\vert V_{3}\right\rangle =\frac{\mathbf{i}%
v_{3}(\lambda_{R}^{2}-\lambda_{L}^{2})}{\sqrt{\lambda_{L}^{2}+\lambda_{R}^{2}%
}}%
\end{equation}
When $\Delta_{K\leftrightarrow K^{\prime}}\neq0$, one has:%
\begin{equation}
\left\langle T_{a}\right\vert h_{so}\left\vert V_{3}\right\rangle
=\frac{\mathbf{i}v_{3}\Delta_{so}^{2}(\lambda_{R}^{2}-\lambda_{L}^{2})}%
{\sqrt{\tilde{\Delta}_{r}\left(  \Delta_{so}^{2}(\lambda_{L}^{2}+\lambda
_{R}^{2})+\frac{\Delta_{K/K^{\prime}}^{2}}{2}(\lambda_{L}-\lambda_{R}%
)^{2}\right)  }}%
\end{equation}
These equations are analogue to Eqs. (\ref{ll}) and (\ref{c2}). In principle,
if the cavity frequency is matching with $E_{3}-\delta\sim2\Delta_{r}+\sqrt
{2}t_{eh}$, there can be lasing transitions from $\left\vert V_{3}%
\right\rangle $ to $\left\vert T_{a}\right\rangle $ since $\left\vert
V_{3}\right\rangle $ is higher in energy than $\left\vert T_{a}\right\rangle
$. The $\left\vert V_{3}\right\rangle \leftrightarrow\left\vert T_{a}%
\right\rangle $ transitions are subradiant since $\left\langle T_{a}%
\right\vert h_{so}\left\vert V_{3}\right\rangle $ cancels for $\lambda
_{L}=\lambda_{R}$. Importantly, $\left\langle T_{a}\right\vert h_{so}%
\left\vert V_{3}\right\rangle $ remains finite for $\Delta_{K\leftrightarrow
K^{\prime}}=0$. Therefore, $\Delta_{K\leftrightarrow K^{\prime}}\neq0$ is not
a fundamental constraint to obtain a subradiant lasing transition in our
system. Nevertheless, in practice, the frequency of the $\left\vert
V_{3}\right\rangle \leftrightarrow\left\vert T_{a}\right\rangle $ transition
is not likely to match with the cavity frequency because $\Delta_{K/K^{\prime
}},\Delta_{so}\gg2\pi\nu_{0}$ is expected \cite{Jespersen,Kuemmeth}. We have
chosen to discuss the lasing transition $\left\vert V_{1}\right\rangle
\leftrightarrow\left\vert T_{-}\right\rangle $ at $\delta\sim2\Delta_{r}$
because it corresponds to a frequency of the order of $\sqrt{2}t_{eh}$, which
is expected to be much smaller \cite{Herrmann:10}.

\section{E. Other possible lasing transitions}

\begin{itemize}
\item In principle, there can be radiative transitions from $\left\vert
T_{-}\right\rangle $ to $\left\vert V_{2}\right\rangle $, which are taken into
account by Eq.(7) of the main text. However, the lasing threshold
corresponding to this transition is not reached in the regime of parameters we
consider, because the population of state $\left\vert T_{-}\right\rangle $ is
negligible for $\delta>2\Delta_{r}$.

\item Since the parameter $\delta$ depends on the dots' gate voltages, cavity
photons can also couple to the CPS through the operator $\hat{\delta}_{diff}=%
{\textstyle\sum\nolimits_{\tau,\sigma,\tau^{\prime},\sigma^{\prime}}}
\left\vert \tau\sigma,\tau^{\prime}\sigma^{\prime}\right\rangle \left\langle
\tau\sigma,\tau^{\prime}\sigma^{\prime}\right\vert $. One finds $\left\langle
V_{1}\right\vert \hat{\delta}_{diff}\left\vert V_{2}\right\rangle =\sqrt
{2}t_{eh}/\sqrt{8t_{eh}^{2}+(\delta-2\Delta_{r})^{2}}$. Hence, in principle,
there can be lasing between the states $\left\vert V_{1}\right\rangle $ and
$\left\vert V_{2}\right\rangle $. Nevertheless, this can be avoided by using
$E_{0}=2\pi\hbar\nu_{0}<2\sqrt{2}t_{eh}$ or by tuning properly $\delta$, so
that $E_{0}\neq E_{1}-E_{2}$.

\item In principle, spin-orbit interaction can also induce lasing transitions
inside the CPS singly occupied charge sector, corresponding to energy
differences $2\Delta_{r}$, $2\Delta_{r}+2t_{ee}$, $2\Delta_{r}-2t_{ee}$ and
$2t_{ee}$. Since the scale $\Delta_{r}$ is expected to be much larger than
$E_{0}$, only the transitions with frequency $t_{ee}/\pi\hbar$ can possibly
match with the cavity frequency $\nu_{0}$. However, it is rather unlikely to
have such a matching in practice. In the main text we assume $\nu_{0}\neq
t_{ee}/\pi\hbar$. This criterion can be checked experimentally by extracting
$t_{ee}$ from the data.
\end{itemize}

\section{F. Evaluation of the spin/photon coupling in a carbon-nanotube based
quantum dot}

In this section, we estimate the spin/photon coupling $\lambda_{L(R)}$ which
can be obtained in the single-wall carbon-nanotube based quantum dot $L(R)$
thanks to spin-orbit coupling.

\subsection{F.1. Electronic wavefunction in the absence of inter-subband
coupling elements}

The position $\overrightarrow{u}$ of an electron on the nanotube is marked
with a longitudinal coordinate $\xi$ and an azimuthal angle $\varphi$, i.e.
$\overrightarrow{u}=\xi\overrightarrow{z}+R\cos[\varphi]\overrightarrow
{x}+R\sin[\varphi]\overrightarrow{y}$ with $R$ the nanotube radius. We write
electronic wavevectors under the form
\begin{equation}
\left\vert \Psi\right\rangle =e^{i\kappa\varphi}\left\vert \psi\right\rangle
\otimes\left\vert \sigma\right\rangle \label{FULL}%
\end{equation}
where $\kappa$ is the electronic circumferential wavevector, $\left\vert
\sigma\right\rangle $ denotes the spin part of the wavefunction and
$\left\vert \psi\right\rangle $ is the $\xi$-dependent orbital part, which has
a structure in sublattice space. We use the spin index $\sigma\in
\{\uparrow,\downarrow\}$ and the sublattice index $\tau\in\{K,K^{\prime}\}$ or
equivalently $\sigma\in\{+,-\}$ and $\tau\in\{+,-\}$ in algebraic expressions.
For a zigzag nanotube, $\left\langle \xi\right.  \left\vert \varphi
\right\rangle $ is an eigenvector of
\[
H_{SWNT}=\hbar v(\tau\kappa s_{1}-is_{2}\frac{\partial}{\partial\xi}%
)+\sigma\tau\Delta_{so}^{0}s_{0}+\sigma\tau\Delta_{so}^{1}s_{1}-\Delta
_{g}s_{1}+V(\xi)s_{0}%
\]
with $V(\xi)$ a longitudinal confinement potential and $\{s_{0},s_{1}%
,s_{2},s_{3})$ the identity and Pauli operators in sublattice space. We have
used the same conventions as in Refs. \cite{Bulaev,Jespersen} to write
$H_{SWNT}$. The motion of electrons along the nanotube circumference is
quantized, i.e.
\[
\kappa=(N+\frac{\tau\eta}{3})\frac{1}{R}%
\]
with $N$ the subband index and $\eta\in\{-1,0,1\}$ a parameter which depends
on the nanotube chiral vector. We have introduced in the above hamiltonian
intra-subband spin-orbit coupling terms in $\Delta_{so}^{1}$, $\Delta_{so}%
^{0}$ and a spin-independent term in $\Delta_{g}$, which are derived e.g. in
Refs. \cite{Izumida,Klinovaja}. The constants $\Delta_{so}^{1}$ and
$\Delta_{so}^{0}$ are first order in the atomic spin-orbit interaction
$V_{so}$ and the nanotube curvature $R^{-1}$, whereas $\Delta_{g}$ is
proportional to $R^{-2}$.

For a problem uniform in the $\xi$ direction ($V(\xi)=0$), the eigenstates
$\left\vert \psi\right\rangle =\left\vert \psi_{\tau,N,k,\sigma,b}%
^{0}\right\rangle $ of the above hamiltonian satisfy $H_{SWNT}\left\vert
\psi_{\tau,N,k,\sigma}^{0}\right\rangle =E_{\tau,N,k,\sigma}\left\vert
\psi_{\tau,N,k,\sigma}^{0}\right\rangle $, with $k$ the electrons longitudinal
wavevector. One can check:
\[
\left\langle \xi\right.  \left\vert \psi_{\tau,N,k,\sigma}^{0}\right\rangle
=\left(
\begin{array}
[c]{c}%
u_{\tau,N,k,\sigma}\\
1
\end{array}
\right)  e^{ik\xi}=\left(
\begin{array}
[c]{c}%
b\frac{\hbar v_{F}\tau\kappa+\sigma\tau\Delta_{so}^{1}-\Delta_{g}-i\hbar
v_{F}k}{\sqrt{\left(  \hbar v_{F}\tau\kappa+\sigma\tau\Delta_{so}^{1}%
-\Delta_{g}\right)  ^{2}+(\hbar v_{F}k)^{2}}}\\
1
\end{array}
\right)  e^{ik\xi}%
\]
and%
\begin{equation}
E_{\tau,N,k,\sigma}=\sigma\tau\Delta_{so}^{0}+b\sqrt{\left(  \hbar v_{F}%
\tau\kappa+\sigma\tau\Delta_{so}^{1}-\Delta_{g}\right)  ^{2}+(\hbar
v_{F}k)^{2}} \label{en}%
\end{equation}
with $b=\pm1$ for the conduction/valence band. In the following, we use $b=1$.
In order to define a quantum dot, we take into account a rectangular
confinement potential
\[
V(\xi)=\left\{
\begin{tabular}
[c]{l}%
$V_{conf}\text{ for }\xi<0$\\
$0\text{ for }0<\xi<L$\\
$V_{conf}\text{ for }\xi>L$%
\end{tabular}
\ \ \right.
\]
We obtain confined electronic states $\left\vert \psi\right\rangle =\left\vert
\Psi_{\tau,N,n,\sigma}\right\rangle $, with $n$ the index corresponding to a
longitudinal confinement of electrons. More precisely, one has
\[
\left\langle \xi\right.  \left\vert \psi_{\tau,N,n,\sigma}\right\rangle
=\left\{
\begin{tabular}
[c]{l}%
$\frac{A_{\tau,N,n,\sigma}}{\sqrt{2\pi}}e^{\tilde{k}_{1}^{n}\xi}\left(
\begin{array}
[c]{c}%
u_{\tau,N,-i\tilde{k}_{1}^{n},\sigma}\\
1
\end{array}
\right)  \text{ for }\xi<0$\\
$\frac{C_{\tau,N,n,\sigma}}{\sqrt{2\pi}}e^{ik_{1}^{n}\xi}\left(
\begin{array}
[c]{c}%
u_{\tau,N,k_{1}^{n},\sigma}\\
1
\end{array}
\right)  +\frac{D_{\tau,N,n,\sigma}}{\sqrt{2\pi}}e^{-ik_{1}^{n}\xi}\left(
\begin{array}
[c]{c}%
u_{\tau,N,-k_{1}^{n},\sigma}\\
1
\end{array}
\right)  \text{ for }0<\xi<L$\\
$\frac{B_{\tau,N,n,\sigma}}{\sqrt{2\pi}}e^{\tilde{k}_{1}^{n}(L-\xi)}\left(
\begin{array}
[c]{c}%
u_{\tau,N,i\tilde{k}_{1}^{n},\sigma}\\
1
\end{array}
\right)  \text{ for }\xi>L$%
\end{tabular}
\ \ \right.
\]
with $k_{1}^{n}>0$ and $\operatorname{Re}[\tilde{k}_{1}^{n}]>0$. The
wavevectors $k_{1}^{n}$ and $\tilde{k}_{1}^{n}$ can be obtained from the
energy-conservation condition
\begin{equation}
E_{\tau,N,k_{1}^{n},\sigma}=E_{\tau,N,-i\tilde{k}_{1}^{n},\sigma}+V_{conf}
\label{Border}%
\end{equation}
the constraint%
\[
n\frac{\pi}{L}<k_{1}^{n}<(n+1)\frac{\pi}{L}%
\]
and the secular condition
\[
\exp\left(  i2k_{1}^{n}L\right)  =\frac{\left(  u_{\tau,N,i\tilde{k}_{1}%
^{n},\sigma}-u_{\tau,N,-k_{1}^{n},\sigma}\right)  \left(  u_{\tau
,N,-i\tilde{k}_{1}^{n},\sigma}-u_{\tau,N,k_{1}^{n},\sigma}\right)  }{\left(
u_{\tau,N,-i\tilde{k}_{1}^{n},\sigma}-u_{\tau,N,-k_{1}^{n},\sigma}\right)
\left(  u_{\tau,N,i\tilde{k}_{1}^{n},\sigma}-u_{\tau,N,k_{1}^{n},\sigma
}\right)  }%
\]
which results from the continuity of $\left\vert \psi_{\tau,N,n,\sigma
}\right\rangle $ at $\xi=0$ and $\xi=L$. The constants $A_{\tau,N,n,\sigma}$,
$B_{\tau,N,n,\sigma}$, $C_{\tau,N,n,\sigma}$, and $D_{\tau,N,n,\sigma}$ can be
obtained from the continuity of $\left\vert \psi_{\tau,N,n,\sigma
}\right\rangle $ and its normalization condition, i.e. $%
{\textstyle\int\nolimits_{-\infty}^{\infty}}
d\xi\left\vert \left\langle \psi_{\tau,N,n,\sigma}\right.  \left\vert
\psi_{\tau,N,n,\sigma}\right\rangle \right\vert ^{2}=1$.

\subsection{F.2. Effect of the inter-subband coupling elements}

In this section, we discuss the coupling between electronic spins and cavity
photons, mediated by the electromagnetic field of the cavity.

\begin{itemize}
\item We assume that the nanotube is parallel to the cavity central conductor.
We take into account the interaction of electrons with the vector potential
$\overrightarrow{A}$ of the cavity treated in the Coulomb gauge
\cite{Bulaev,CT}. We quantize the field $\overrightarrow{A}$ in terms of the
photonic operators\cite{Gardiner,Blais}. This gives a coupling operator%
\[
H_{inter}^{A}=-\frac{e\hbar V_{rms}}{8\pi m_{eff}R\nu_{0}d}(a+a^{\dag}%
)[(\mu_{+}-\mu_{-})\frac{\partial}{\partial\varphi}+\frac{\partial}%
{\partial\varphi}(\mu_{+}-\mu_{-})]
\]
We have used above $\sin(\varphi)=(\mu_{+}-\mu_{-})/2i$, where the operator
$\mu_{\pm}=e^{\pm i\varphi}$ increases/decreases the index $N$.

\item The subband and spin subspaces are coupled by a term
\cite{Izumida,Klinovaja}
\[
H_{inter}^{so}=-\Delta_{so}^{1}s_{2}i(\sigma_{-}\mu_{+}-\sigma_{+}\mu_{-})
\]
which is first order in the atomic spin-orbit interaction $V_{so}$ and the
nanotube curvature $R^{-1}$ (spin-orbit interaction in carbon nanotubes was
also discussed in Refs. \onlinecite{Huertas,Ando,DeMartino,Jeong,Bulaev}).
Here, $\sigma_{\pm}$ is the operator increasing/decreasing the spin index in
$\left\vert \sigma\right\rangle $.

\item Note that the terms $H_{inter}^{A}$ and $H_{inter}^{so}$ apply to the
full wavefunctions [see Eq. (\ref{FULL})]
\[
\left\vert \Psi_{\tau,N,n,\sigma}\right\rangle =e^{i(N+\frac{\tau\eta}%
{3})\varphi}\left\vert \psi_{\tau,N,n,\sigma}\right\rangle \otimes\left\vert
\sigma\right\rangle
\]
One can perform a first order perturbation of these wavefunctions by
$H_{inter}^{so}$:
\[
\left\vert \tilde{\Psi}_{\tau,N,n,\sigma}\right\rangle =\left\vert \Psi
_{\tau,N,n,\sigma}\right\rangle +%
{\displaystyle\sum\limits_{n^{\prime}}}
\lambda_{\tau,N,n,n^{\prime}}^{-\sigma}\left\vert \Psi_{\tau,N+\sigma
,n^{\prime},-\sigma}\right\rangle
\]
with%
\[
\lambda_{\eta,\tau,N,n,n^{\prime}}^{\sigma}=\frac{\left\langle \Psi
_{\tau,N-\sigma,n^{\prime},\sigma}\right\vert H_{inter}^{so}\left\vert
\Psi_{\tau,N,n,-\sigma}\right\rangle }{E_{\tau,N,n,-\sigma}-E_{\tau
,N-\sigma,n^{\prime},\sigma}}%
\]

\item The term $H_{inter}^{A}$ couples the perturbed wavefunctions $\left\vert
\tilde{\Psi}_{\tau,N,n,\sigma}\right\rangle $ for $\sigma=\uparrow$ and
$\sigma=\downarrow$. This yields intra-subband spin/photon coupling elements,
which write for the lower subband $N=0$ which we consider:
\begin{align*}
\lambda_{\tau}  &  =\left\langle \tilde{\Psi}_{\tau,N=0,n,-1}(\overrightarrow
{r})\right\vert H_{inter}^{A}\left\vert \tilde{\Psi}_{\tau,N=0,n,+1}%
(\overrightarrow{r})\right\rangle \\
&  =P\left(  \frac{\tau\eta}{3}\left(  \lambda_{\tau,N=0,n,n}^{-}+\left.
\lambda_{\tau,N=0,n,n}^{+}\right.  ^{\ast}\right)  +\frac{1}{2}\left(
\lambda_{\tau,N=0,n,n}^{-}-\left.  \lambda_{\tau,N=0,n,n}^{+}\right.  ^{\ast
}\right)  \right)
\end{align*}
with $P=i e\hbar V_{rms}/8\pi m_{eff}R\nu_{0}d.$

\item One can check that $\lambda_{K}=\lambda_{K^{\prime}}=-i\lambda$ is
purely imaginary due to the assumptions used above [zigzag nanotube and
$V(\xi-\frac{L}{2})=V(\frac{L}{2}-\xi)$]. One thus obtains the intra-subband
spin/photon coupling term of Eq.(3) of the main text. For $L=100~\mathrm{nm}$,
$R=1~\mathrm{nm}$, $\eta=1$, $d=5~\mathrm{\mu m},$ $V_{rms}=4~\mathrm{\mu V}$,
$\nu_{0}=3.64~\mathrm{GHz}$, $E_{conf}=\hbar v_{F}/3R$, $m_{eff}%
=\hbar\left\vert \kappa\right\vert /v_{F}=4.4$ $10^{-32}~\mathrm{kg}$, and
using the parameters $\Delta_{so}^{1}=-0.08~\mathrm{meV}$, $\Delta_{so}%
^{0}=-0.32~\mathrm{meV}$, and $\Delta_{g}=5.7~\mathrm{meV}$ taken from Ref.
\onlinecite{Klinovaja}, we obtain $\lambda\simeq0.4~\mathrm{MHz}$.

\item In order to obtain a tunable spin/photon coupling, one can insert in one
of the dots gate voltage supply a tunable capacitance made out of a single
electron transistor (SET). The electric field seen by the nanotube can be
modulated electrostatically by placing the SET in the blockaded or
transporting regimes. This allows one to vary the couplings $\lambda_{L(R)}$
for dot $L(R)$.

\item In the case $\lambda_{iK}\neq\lambda_{iK^{\prime}}$, with ${i}$ the dot
index added in the main text, the results of the main text can be generalized
straightforwardly by using
\begin{equation}
\left\langle T_{_{-}}\right\vert h_{so}\left\vert V_{n}\right\rangle
=v_{n}\frac{\Delta_{K\leftrightarrow K^{\prime}}}{2\Delta_{r}}(\lambda
_{LK}+\lambda_{LK^{\prime}}-\lambda_{RK}-\lambda_{RK^{\prime}})
\end{equation}

\end{itemize}

\end{document}